# Everettian theory as pure wave mechanics plus a no-collapse probability postulate

Paul Tappenden  paulpagetappenden@gmail.com

> May the spirit of Newton's method give us the power to restore unison between physical reality and the profoundest characteristic of Newton's teaching – strict causality.
>
> (Einstein 1927, p. 467)

> The theory based on pure wave mechanics is a conceptually simple, causal theory.
>
> (Everett 1957, p. 462)

**Abstract**

Proposed derivations of the Born rule for Everettian theory are controversial. I argue that they are unnecessary but may provide justification for a simplified version of the Principal Principle. It's also unnecessary to replace Everett's idea that a subject splits in measurement contexts with the idea that subjects have linear histories which partition (Deutsch 1985, 2011; Saunders and Wallace 2008; Saunders 2010; Wallace 2012, Chapter 7; Wilson 2013; forthcoming). Linear histories were introduced to provide a concept of pre-measurement uncertainty and I explain why pre-measurement uncertainty for splitting subjects is after all coherent, though not necessary because Everett's original *fission* interpretation of branching can arguably be rendered coherent without it, via reference to (Vaidman 1998; Tappenden 2011; Sebens and Carroll 2018; McQueen and Vaidman 2019). A deterministic and probabilistic quantum mechanics can be made intelligible by replacing the standard collapse postulate with a no-collapse postulate which identifies objective probability with relative branch weight, supplemented by the simplified Principal Principle and some revisionary metaphysics.

**1 Everett's sputnik**

In 1957 Hugh Everett III launched what he called 'a conceptually simple, causal theory' which he described as 'based on pure wave mechanics' (Everett 1957, p. 462). More than sixty years on the idea is still up there and much discussed, yet scholars remain divided as to how to make sense of it and even whether it can possibly make any sense at all. So Everett's idea appears not to be so conceptually simple as he thought, or at least not obviously so.

Jeffrey Barrett has cogently argued that whilst Everett's theory may be *based* on pure wave mechanics it cannot be reduced to wave mechanics alone (Barrett 2017, p. 31). Everett sought to eliminate the collapse postulate from standard quantum theory because it effectively adds a putative stochastic process to the purely deterministic mechanism of unitary evolution. The collapse postulate, via the Born rule, is often construed as assigning objective probabilities to what are understood to be mutually exclusive 'possible' outcomes of quantum measurement events, nowadays thought of more broadly as types of event involving decoherence. As Barrett stresses, Everett understood eliminating the collapse postulate to entail eliminating the concept of objective probability itself and sought to ground statistical evidence on a notion of typicality (*ibid.*, p. 33).

Here Everett arguably went astray, though understandably so, since replacing the collapse postulate with a no-collapse postulate is highly counterintuitive. All that follows is aimed at overcoming those negative intuitions. Rather than interpreting the Born rule as assigning objective probabilities to possible future outcomes of measurement-like events, a no-collapse postulate should interpret it as assigning objective probabilities to future co-existent actual outcomes, which Everett called 'branches'.

If objective probability is thought of in this way it becomes a relation involving the absolute square of the quantum amplitude of each branch, a quantity which has come to be known as branch 'weight'. The objective probability of a downstream branch is its weight relative to the weight of the upstream branch from which it emanates. This idea is not new. It has been expressed very clearly by Simon Saunders (1993, §7) and by David Papineau (1995, 2010). But it has been widely thought that something more is needed than to simply replace the standard collapse postulate by the no-collapse postulate. For instance Saunders, and others, have proposed ways to deal with an apparent lack of pre-measurement uncertainty in Everettian theory, of which more later. And there are several rather complex arguments which purport to show that the subjective probabilities assigned to future observations of quantum measurement outcomes should be numerically equal to the relative weights of the branches where those observations will be made. A result which can be interpreted as suggesting that the weights of those branches might just *be* their objective probabilities relative to the ready state of the measurement device. But those arguments are controversial.

However, just as we are free to *posit* objective probability as associated with so-called possibilities (or propensities) for standard stochastic quantum mechanics, so we are free to *posit* the identification of relative branch weight with objective probability for a non-standard 'dendritic' quantum mechanics. In both cases the posit is aimed at justifying a subject's assignment of *subjective* probabilities to the observation of distinct futures arising from a set of initial conditions. A posit is simply a hypothesis and the hypothesis that relative branch weight is objective probability can provide the basis for a probabilistic *and* deterministic wave mechanics.

If objective probability does indeed exist out there in the quantum world then that's a novel discovery which the study of radiation and particles has led to. In that case objective probability is a something-we-know-not-what whose existence gives rise to the observation of frequency distributions. We have no reason to suppose that our ordinary uses of the term 'probability' should in any way determine the nature of that something. Everettian theory arguably suggests that that something just is relative branch weight. As such, the nature of branch weight would seem to be that of a sort of non-spatiotemporal extension. Neither a length nor a duration, but something of that ilk. As if a branch were like a road of width X and its bifurcation like that of a road dividing into roads of widths Y and Z, where Y+Z=X. The idea is reminiscent of what Michael Lockwood called a superpositional 'dimension', attributing the idea to David Deutsch (Lockwood 1989, p. 232).

  Preposterous as this may seem, that is the idea which I shall be defending in what follows. The proposal is that replacing the conventional collapse postulate of quantum mechanics with the no-collapse postulate can provide the basis for making Everettian probability fully intelligible, something which remains a matter of deep dispute amongst experts. In order to explain how that can be so I shall begin by considering the ways in which the concepts of branching and uncertainty have arisen in Everettian theory.

## 2 Quantum fission

Everett suggested that for an idealized experimental setup, where the spin of an x-spin-up particle is measured on a different axis, the apparatus 'splits' onto branches where the apparatuses on each branch are the elements of an apparatus in superposition on the pointer basis, with an *up* element indicating spin-up and a *down* element indicating spin-down[1]. Objects in the environment of the apparatus become entangled with it and are thereby also caused to evolve into superpositions and, in particular, the body of any subject observing the apparatus is caused to split and the elements of that corporeal superposition register the observation of *up* and of *down*[2]. Everett wrote:

> The 'trajectory' of the memory configuration of an observer performing a sequence of measurements is thus not a linear

---

[1] The pointer basis is arguably determined by the process of decoherence. See (Wallace 2012, Ch. 3) for details. The measurement idealization ignores many bizarre low amplitude branches and so takes the absolute square of amplitude for the *up* and *down* branches to sum to 1.

[2] Roland Fraïssé argued that a splitting event gives rise to a *ramifier*, the lightspeed propagation of the branching of spacetime (Fraïssé 1974; 1982; 1986). Given that any splitting event creates a gravitational disturbance, this suggests an interface between general relativity and quantum mechanics. Note that the different views in (McQueen and Vaidman 2019) and (Sebens and Carroll 2018) turn on a difference in their interpretations of the process of branching.

> sequence of memory configurations, but a branching tree, with all possible outcomes existing simultaneously.
>
> (*op.cit.*, p. 460)

In an unpublished manuscript he also wrote:

> As an analogy one can imagine an intelligent amoeba with a good memory. As time progresses the amoeba is constantly splitting, each time the resulting amoebas having the same memories as the parent. Our amoeba hence does not have a life line, but a life tree.
>
> (Barrett and Byrne 2012, p. 69)

This concept of dendritic structure needs some unpacking for there seem to be two distinct types of 'splitting' involved. When an amoeba splits we would normally say that the two downstream cells are parts of the original. Similarly the *up* and *down* elements of the superposed apparatus can be thought of as 'superpositional' parts of a single object, an idea which David Wallace refers to as the *Hydra View* (*op.cit.*, p. 281). Since the measurement process involves decoherence, the *up* and *down* branches of the apparatus are effectively causally isolated from one another.

But when Alice makes the idealized spin measurement she splits into Alice$_{up}$ and Alice$_{down}$ who observe the outcomes *up* and *down*. And we think of Alice$_{up}$ and Alice$_{down}$ as distinct subjects, not as parts of a single subject. This is not amoeba-like splitting, it's personal fission, a topic which has been much discussed independently of Everettian theory. Note that Everett attributes a dendritic structure to 'memory configurations' and that concept can indeed be subsumed under the Hydra view. A recording device coupled with the measurement apparatus will evolve into a superposition whose elements are devices recording *up* and *down* and those elements can be considered as superpositional parts of a single superposed recording device.

So it appears to be the concept of subjects, not that of objects, which is what creates problems for Everett's allusion to amoeba-like splitting. We cannot conceive of a single subject observing *up* and *down* simultaneously. At the same time, it does seem that *brains* can be thought of as recording devices. Alice's brain prior to measurement, like the apparatus, can be understood to evolve into a superposition with two elements, a brain recording *up* and a brain recording *down*, which, as before, can be regarded as parts of a single superposed brain. This can seem worryingly mysterious, but it need not. For consider an inscription of the sentence 'the result of experiment X is *up* and the result of experiment X is *down*'. The inscription contains the inscriptions of two sentences as parts whilst the contained inscriptions are of sentences which express contradictory propositions. As will become clear, making the no-collapse postulate intelligible apparently requires brains to have a similar sort of relation to observations as do inscriptions to sentences, a variety of type-token relation. So whilst Alice$_{up}$ and Alice$_{down}$ are distinct subjects making contradictory observations, their brains can be parts of a single cerebral superposition.

But if Alice$_{up}$ and Alice$_{down}$ are distinct subjects they cannot *both* be identical to Alice. And there seem to be no grounds for believing that Alice has become one or the other of them. Some Everettian theorists have glossed over this difficulty by simply referring to Alice$_{up}$ and Alice$_{down}$ as Alice's 'successors' (Papineau 2003) or 'descendants' (McQueen and Vaidman 2019). The problem appears to be this. On a standard stochastic interpretation of the measurement setup Alice will either observe *up* or observe *down*. She'll have a single 'descendant', herself. So it would seem that she can unproblematically be uncertain about what she'll observe. She can assign *subjective* probabilities to the future observation of *up* or *down* on the basis of her assessment of the *objective* probabilities for each possibility. But in the face of personal fission how can Alice even expect to *survive*?

I shall not attempt here to survey the extensive literature on personal fission but shall adopt what strikes me as the most perspicuous way of dealing with the problem, Ted Sider's *stage theory* (Sider 1996; 2001, p. 201)[3]. Others may prefer a different analysis of trans-temporal identity. A useful discussion of some alternatives and their comparison with stage theory can be found in (Hawley 2001).

**2.1 Stage theory**

Stage theory is inspired by modal counterpart theory where a subject or object is not numerically identical with the modal counterpart which it might have been. Similarly, subjects and objects are not taken to be numerically identical with what they were and what they will be. So Alice$_{up}$ and Alice$_{down}$ are what have come to be known as *future temporal counterparts* of Alice, which entails that she bears the relation *will be* to Alice$_{up}$ and to Alice$_{down}$. Contrariwise, Alice is a *past temporal counterpart* of both Alice$_{up}$ and Alice$_{down}$ so each of them bears the relation *was* to Alice.

Stage theory certainly has some counterintuitive consequences, as is well recognized. For instance, a person in prison is never the same person as the one who committed the crime. However, if the person in prison *was* the person who committed the crime that could seem a good enough justification for the person in prison *being* in prison. Also, if you've apparently been alone in a room for an hour stage theory has it that there have been many momentary persons in many momentary rooms during that hour. Still, at any given moment there's only one person in one room and each of those persons, except the first one, *was* each of the earlier persons.

But there remains a problem for applying stage theory to Everettian personal fission which has not, so far as I know, been addressed. In a context such as that of our model spin measurement the temporal counterpart relations for subjects and objects need to be different. The immediate post-measurement temporal counterpart of Alice's brain includes two brains which are superpositional parts of a single object, one part recording *up* and the other part recording *down*. The superposition of the two brains is a single future temporal counterpart of Alice's brain. Her brain evolves into a superposition. It's spatiotemporal continuity which connects Alice's brain with its future temporal counterparts. But the *pair* [Alice$_{up}$ and Alice$_{down}$] had better not be a future temporal counterpart of Alice for reasons which will become apparent. Resolving this problem requires further work which I'll come to in §2.4.

---

[3] Hilary Greaves introduced stage theory independently to Everettian theory, without reference to Sider (Greaves 2004, §4.1.1). Without reference to Greaves, but with reference to Sider, stage theory is employed in an Everettian context in (Tappenden 2008, p. 313).

But even given a resolution of this problem for stage theory there's a further problem for fission. Alice, if well informed, knows that she'll split. She will be Alice$_{up}$ and she will be Alice$_{down}$, though she will not become a pair of people. But in what sense is she *uncertain* about her future experience?

## 2.2 Lewisian uncertainty debugged

If Alice believes that she'll split it can seem obvious that she must lack uncertainty about the outcome of her measurement. But if she lacks uncertainty how can she intelligibly assign *probabilities* to outcomes? Hilary Greaves refers to that as the *incoherence problem* for Everettian theory (Greaves 2004, §1). Her proposed solution to the incoherence problem involved rejecting the need for uncertainty. She contrasted fission with what she called the *subjective uncertainty* view for which Saunders and Wallace were later to become champions. The subjective uncertainty program has sought pre-measurement uncertainty by replacing Everett's fission model of branching with what can be called the *partitioning linear histories* model, which I'll discuss in §3.

For standard quantum theory uncertainty as to future observations consists in assigning subjective probabilities to those futures and for a stochastic theory which associates objective probabilities with those futures subjective probabilities are assigned on the basis of what David Lewis has called the Principal Principle (Lewis 1980, p. 266). Since the no-collapse postulate also associates objective probabilities with measurement outcomes any appeal to uncertainty must involve something like the Principal Principle though it turns out to differ somewhat from what Lewis had in mind.

Lewis's concept of objective probability was stochastic. That's clear when he writes:

> Next question. As before, except that now it is afternoon and you have evidence that became available after the coin was tossed at noon. Maybe you know for certain that it fell heads; maybe some fairly reliable witness has told you that it fell heads; maybe the witness has told you that it fell heads in nine out of ten tosses of which the noon toss was one. You remain as sure as ever that the chance of heads, just before noon, was 50%. To what degree should you believe that the coin tossed at noon fell heads?
> Answer. Not 50%, but something not far short of 100%.
> 
> (*ibid.,* p. 265)

Lewis's coin stands in for stochastic quantum processes; he earlier makes reference to the decay of a tritium nucleus. Just as the nucleus does or does not decay within a given period, so the coin falls exclusively either heads or tails. Lewis concludes as follows:

> If evidence bears in a direct enough way on the outcome - a way that may nevertheless fall short of outright implication - then it may bear on your beliefs about outcomes otherwise than by way of your beliefs about the chances of the outcomes. Resiliency under all evidence whatever would be extremely unreasonable. We can only say that degrees of belief about outcomes that are based on certainty about chances are resilient under *admissible* evidence. The previous question gave examples of admissible evidence; this question gave examples of inadmissible evidence.
>
> (*ibid*., original emphasis)

Inadmissible evidence for Lewis is evidence about the future. If you believe that the chance that the coin will fall heads at noon is 50% but 'some fairly reliable witness' tells you that after noon that the coin has fallen heads, but you have not been able to see the result for yourself, then your belief that the coin has fallen heads becomes 'something not far short of 100%'.

As Mauricio Suárez puts it:

> On the Humean approach defended by David Lewis, for example, chance is a function of the entire state or history of the world up to a certain time, that is, $ChHt(x)$, where Ht is the history of the world, w, up to time t. Chance is thus both world and time relative (Lewis 1986, p. 91).
>
> (Suárez 2017, p. 1167)

Suárez's assertion that Lewisian chance is world-relative refers to Lewis's 'modal realism', fully presented in (Lewis 1986). If Lewis's concept of a stochastic process is replaced by that of a branching process the idea of inadmissible evidence disappears. Wallace has made this point (*op. cit.*, p. 150). Here I expand on the idea in the context of viewing Everettian subjects as having dendritic histories which, as we'll see later, Wallace rejects. Post-measurement, Alice$_{up}$'s observation of *up* doesn't confirm that the apparatus which Alice faced in the ready state has come to indicate *up*. Likewise for Alice$_{down}$. Because what Alice$_{up}$ and Alice$_{down}$ each identify as a measuring device is but a superpositional element of the device into which the original has evolved. Given the no-collapse postulate, if Alice were to have complete knowledge of the future it would tell her that her apparatus would fission into an apparatus indicating *up* on a branch who's objective probability is, say, 0.7 and an apparatus indicating *down* on a branch who's objective probability is 0.3. And that knowledge would do nothing to undermine her prior judgment of the objective probabilities and so there would be no reason to count it inadmissible.

Lewis was viscerally opposed to the idea of what he called a branching world, writing:

> The trouble with branching exactly is that it conflicts with our ordinary presupposition that we have a single future. If two futures are equally mine, one with a sea fight tomorrow and one without, it is nonsense to wonder which way it will be – it will be both ways – and yet I do wonder. The theory of branching suits those who think this wondering *is* nonsense.
>
> (Lewis 1986, p. 209, original emphasis)

Quite rightly, Lewis sees the idea that a person can fission as unacceptable because 'it conflicts with our ordinary presupposition that we have a single future'. And he specifically links that opposition with rejection of Everettian theory in a letter dated 21st December 1987 (Beebee and Fisher, forthcoming). In what follows I shall explain how 'wondering' about quantum measurement outcomes need *not* be nonsense in the absence of stochasticity. From the dendritic point of view Lewis's Principal Principle doesn't need to be qualified in order to take into account the admissibility of evidence, it can simply be taken as stating that subjective probabilities should be assigned equal to what the relevant objective probabilities are believed to be. Of course judgment as to what the value of relevant objective probabilities *are* will be based on statistical evidence, but that's a different issue which I'll come to in §4.

But the simplified Principal Principle which I've just described, and shall hereafter refer to as PP, is an *assumption*. It has an intuitive appeal; it seems obvious that if you believe that the objective probability of some event occurring is X then you should assign a subjective probability of X to the future observation of that event, other things being equal. But I shall be arguing that some claimed derivations of the Born rule for Everettian theory may be reinterpreted as *justifications* of PP. In the meantime, the combination of the no-collapse postulate and PP entails that Alice pre-measurement assigns subjective probabilities to multiple co-existent futures.

**2.3 Fission and pre-measurement uncertainty**

Tim Maudlin has written:

> It is easy to state the problem of probability in the Everett theory: probabilities are standardly attached to *alternatives*
>
> (Maudlin 2014, pp. 799-800, original emphasis)

If Alice believes that she will split and assigns subjective probabilities of 0.7 and 0.3 to her future observation of *up* and of *down* no alternatives seem to be involved so if Maudlin's point is to be met some sort of *no-standard* concept is called for. I have introduced a non-standard concept of the mind-body relation which appears to allow probabilities to attach to co-existents rather than alternatives (Tappenden 2017, §2).

Consider a large but not infinite set of isomorphic universes in which quantum measurement processes are stochastic. At corresponding spactime locations in each universe a version of our model spin measurement takes place with the result that, given the law of large numbers, the original set of universes partitions into a subset of measure 0.7 where *up* occurs and a subset of measure 0.3 where *down* occurs.

Now introduce subjects. The standard interpretation of the setup would introduce corresponding subjects, one in each universe. Each would be manifest as a doppelganger beside the measurement apparatus. And it could be supposed that each subject believes that the measurement process about to take place is stochastic and that the objective probabilities for the outcomes *up* and *down* are 0.7 and 0.3 respectively. That belief might have arisen from each subject having done a series of tests with the measuring apparatus.

A non-standard way to interpret the setup is to suppose that there's a *single* subject whose body is the *set* of doppelgangers. What that observer refers to as a single measuring apparatus is the *set* of corresponding apparatuses, one in each universe. That is what I have called the *unitary interpretation of mind*. On this analysis both the 'unitary' subject and the 'plural' subjects are in *exactly the same mental state*. All the difference is in the worlds, not in the minds of the subjects. What has changed is that there is one *token* of that mental state rather than multiple tokens and it's integral to the alternative interpretation that the constitution of objects in the subjects' environments changes too. There's more on that in §5. The unitary subject has just the same beliefs as the original multiple subjects and so believes that the measuring device to which s/he refers will exclusively yield the result *up* or *down* and s/he assigns objective probabilities of 0.7 and 0.3 to those alternatives.

But the unitary subject is *mistaken*. What *in fact* happens to the single measuring device is that it *splits* into two devices, each of which is a subset of the original. And the subject also splits into two subjects, one observing *up* and the other observing *down*. Each subject's body is a set of doppelgangers in a subset of the original set of universes, the subset measures being 0.7 and 0.3.

What the thought experiment shows is that if the unitary interpretation of mind is coherent then it's coherent that a subject who believes s/he faces a stochastic process in fact faces a dendritic process. So it's coherent that an assignment of objective probabilities to what are believed to be *alternative* outcomes is in fact an assignment to *co-existent* outcomes. The question which then arises is what difference it makes if the subject comes to *believe* that s/he faces a dendritic process rather than a stochastic process. Does s/he cease to be *uncertain* about what will happen?

Alice knows that she'll split into Alice$_{up}$ observing *up* on a branch of weight 0.7 and Alice$_{down}$ observing *down* on a branch of weight 0.3. She also knows that the outcome *up* will occur and that the outcome *down* will occur. If the outcome *up* is going to occur the objective probability that it will occur is 1, no?

That's an intuition which can be resisted. The objective probability of the occurrence of *up* is 0.7 because that's the objective probability of the branch on which it occurs relative to the ready state's branch, according to the no-collapse postulate. What has objective probability 1 is the combined occurrence of both *up* and *down* since 0.7 + 0.3 = 1.

If that's right it's possible to understand Alice as being *uncertain* as to what she'll observe because, via the no-collapse postulate and PP, she assigns subjective probabilities of 0.7 to her future observation of *up* and 0.3 to her future observation of *down*. She does so because, applying stage theory, she knows she will be Alice$_{up}$ on a branch whose objective probability is 0.7 and she will be Alice$_{down}$ on a branch whose objective probability is 0.3. What Alice is *certain* of is that *both* outcomes will occur but she doesn't assign a subjective probability of 1 to observing both outcomes because she does not believe that she will be the pair [Alice$_{up}$ and Alice$_{down}$]. She's not uncertain about *which* outcome will occur but she *is* uncertain about what she'll observe.

That's a response to the incoherence problem for Everettian theory. But its intelligibility seems to depend on a non-standard and counterintuitive interpretation of the mind-body relation. What I shall do now is consider more conservative responses to the incoherence problem and assess their relation to the unitary interpretation of mind. But first a further note on stage theory in the light of this section.

## 2.4 Stage theory again

Recall that for stage theory to be coherently applied to quantum fission there must be different temporal counterpart relations for subjects and objects. Alice's brain's future temporal counterpart is a pair of brains, each a superpositional part of a superposed brain. Alice herself cannot have the pair [Alice$_{up}$ and Alice$_{down}$] as a future temporal counterpart because that would mean that Alice will become a pair of people having contradictory experiences, which is hard to make sense of at the very least. The unitary interpretation of mind makes it possible to sidestep this problem.

The reason is that the criteria for individuating subjects and objects become different. Whereas objects such as brains are individuated by their spatiotemporal location, subjects are individuated by their cognitive contents. The connection which makes Alice$_{up}$ and Alice$_{down}$ future temporal counterparts of Alice is cognitive continuity, not spatiotemporal continuity. Alice$_{up}$ and Alice$_{down}$ are distinct individual subjects who have all their cognitive contents in common, except for their perceptions of the outcomes of Alice's measurement. So whereas Alice's brain's future temporal counterpart is a single superposition with brains as elements, Alice's future temporal counterparts do not include a pair of subjects observing both *up* and *down*. There's just Alice$_{up}$ individuated by her cognitive content and Alice$_{down}$ individualted by *her* cognitive content.

Subsequently, Alice$_{up}$ and Alice$_{down}$ can become much more different cognitively, and go on to fission in different ways. The unitary interpretation of mind entails that two or more brains (or functionally equivalent objects) which instance the same cognitive content, however distributed in spacetime, instance a single individual subject. If, as a result of a climax of improbability, a brain were to come into existence in the Andromeda galaxy which was isomorphic to Alice$_{up}$'s brain, that distant object would also instance Alice$_{up}$'s mind, despite having no spatiotemporal connection with Alice's brain. Bizarre as it is, that's the proposal. And there's no implication of mind-body dualism, as traditionally understood. As I said earlier, it's rather like the relationship between a sentence and its inscriptions. Except that for the putting of marks on paper to be a token inscription it must be causally, and so spatiotemporally, connected with the 'brain' of a subject.

Everett can reasonably be said to have replaced the conventional stochastic interpretation of quantum mechanics with a dendritic interpretation, where subjects and objects 'split' in measurement contexts. But in combining stage theory with Everettian theory it's important to note that the histories of subjects and objects are dendritic in different ways. Objects branch in a sense similar to the way a river branches into an estuary: just as all the branches are parts of the same river, so the future branches of Alice's brain are superpositional parts of one superposed brain. The history of Alice's brain is *linear*, its temporal parts are well ordered. But $Alice_{up}$ and $Alice_{down}$ are not parts of one subject, they're simultaneous temporal parts of a single history whose temporal parts are *partially* ordered, not totally ordered. An advantage of stage theory is that it can describe both histories which have totally ordered temporal parts and histories which have partially ordered temporal parts. However, for stage theory to be employed in describing the histories of Everett's fissioning subjects the unitary interpretation of mind is apparently required in order to distinguish between the temporal counterpart relations for subjects and objects.

**2.5 Greaves and Vaidman on incoherence**

I shall now set aside the previous argument for the coherence of pre-measurement uncertainty for the fission model of branching and consider what other responses to the incoherence problem have been proposed by fission theorists. We shall see that the arguments lead us back to the unitary interpretation of mind.

Greaves responded to the problem by dismissing any need for uncertainty and suggesting that Alice can 'care' about future outcomes, developing the idea of what she called a *caring measure* (*op. cit.*, §2.2). In a later paper she acknowledges that a similar idea had earlier been proposed by Lev Vaidman (2002, §6.4). The way the Vaidman-Greaves caring measure is meant to deal with the incoherence problem is by providing reasons for a subject to act pre-measurement in a similar way to the way s/he would act if believing a stochastic interpretation of the measurement process. That is, the caring measure is supposed to induce a subject to act *as if* the outcomes were alternative possibilities with associated objective probabilities (Greaves 2004, §1).

The paradigm testbed for this idea is a gambling setup where Alice is presumed to be a betting woman and is offered wagers on outcomes. According to Vaidman and Greaves, Alice should care about rewards and losses as a function of their perceived utility combined with the weights of the branches on which they occur. With the result that Alice places bets in what she believes to be a dendritic setup in the same way as she would have done if she had believed it was a stochastic setup. Greaves concludes:

> I have argued (section 2) that the Everettian has no need to claim title to the term 'probability', over and above her needs (a) to formulate a strategy for rational action in the face of branching, and (b) to be entitled to regard quantum mechanics, given the sequences of experimental outcomes we have in fact observed, as empirically confirmed.
>
> (*ibid.*, §6)

Here (a) is what Greaves calls the *practical problem* for Everettian theory and (b) is the *epistemic problem,* which I shall discuss in §4. It's in response to the practical problem that Greaves and Vaidman invoke the concept of caring measure. The idea has been criticized at length by David Albert who forcefully concludes that the strategy 'looks silly and sneaky and unmotivated and wrong' (Albert 2010, p.364).

However, in thinking that Everettian fission requires abandoning uncertainty Greaves overlooked something, as did Saunders when he subsequently characterized fission thus:

> The attempt to ground EQM [Everettian Quantum Mechanics] on [statistics and rationality] alone, disavowing all talk of probability and uncertainty, has been dubbed the *fission programme.*
>
> (Saunders 2010, p. 183)

What both Greaves and Saunders overlook in their characterizations of fission is Vaidman's introduction of the concept of post-measurement, pre-observation uncertainty, of which more shortly (Vaidman 1998, p. 253). Greaves at least acknowledges it but sees the concept of caring measure as having explanatory priority (*op.cit.*, §4.3). I shall now argue that, on the contrary, it is post-measurement, pre-observation uncertainty which explains pre-measurement caring, thereby providing a response to Albert.

Vaidman's idea is this. If $Alice_{up}$ and $Alice_{down}$ are 'blindfolded', i.e. cognitively isolated from the pointer readings, each can be uncertain as to which branch she's located on post-measurement. Each doesn't know whether she's on the *up* branch or the *down* branch. The *assumption* that blindfolded $Alice_{up}$ and $Alice_{down}$ should assign subjective probabilities equal to the branch weights has been called the *Born-Vaidman rule* (Tappenden 2011, §2). It requires justification, but even given that assumption there have been some dismissive responses to Vaidman's idea. Albert has written:

> The trouble with Lev's uncertainty is that it seems altogether avoidable, and that it comes too late in the game. The uncertainty we *need* – the uncertainty that quantum mechanics imposes on us – is something not to be bypassed
>
> (*op.cit.,* pp. 367-8, original emphasis)

In response to Albert, I have argued that something which cannot be bypassed is the *possibility* of Vaidman's post-measurement, pre-observation uncertainty (Tappenden 2011, §4). Prior to placing a bet, Alice knows that it would be possible for her to be ignorant of the outcome post-measurement. All she'd need to do is wear a blindfold. And in such a state of ignorance Alice can be confident that she would make exactly the *same* betting judgment if she could place a stake before removing the blindfold. But the rule is that stakes must be laid *before* the measurement. Knowing that, Alice in a state of Vaidmanian ignorance would *regret* not having laid a stake if she'd not done so. Alice, knowing in advance that she would regret not having laid a stake if she were in a state of Vaidmanian ignorance post-measurement has every reason to lay that stake pre-measurement.

A complete rationale for pre-measurement decision-making is provided by the *possibility* of Vaidmanian ignorance. And the Vaidman-Greaves concept of caring measure is effectively transformed so that what Alice needs to care about pre-measurement it the betting behavior that she would regret not having executed if she were in a state of Vaidmanian ignorance. The unbypassable *possibility* of Vaidmanian ignorance does the job just as well as pre-measurement ignorance, and on reflection that ought not to be surprising. Think about betting on ordinary dice rolls. The way you bet is just the same whether your eyes are open prior to the roll or closed after the roll, given that the payouts on different numbers remain the same. The rule that bets must be laid before the roll is made for the sighted. However, the argument does depend on justifying the Born-Vaidman rule.

**2.6 Justifying the Born-Vaidman rule**

Returning to our model spin measurement, it could seem that Alice$_{up}$ and Alice$_{down}$ in a state of Vaidmanian ignorance must be constrained by a principle of indifference. Each Alice knows that she's one of two people and each, for all she knows, could be either one. So each should assign subjective probabilities of 0.5 to being on the *up* branch and to being on the *down* branch, *irrespective of the branch weights*. Clearly the Everettian project would be undermined in one fell swoop if blindfolded Alice$_{up}$ and Alice$_{down}$ do not have reason to assign subjective probabilities to being on the *up* or the *down* branch equal to the branch weights.

The task of justifying the Born-Vaidman rule has been addressed in (Sebens and Carroll 2018) and (McQueen and Vaidman 2019). The arguments differ in their interpretation of the branching process but both conclude that blindfolded Alice$_{up}$ and Alice$_{down}$ should indeed assign subjective probabilities of 0.7 and 0.3 to the future observation of *up* and *down* respectively, counterintuitive as that may seem.

Sebens and Carroll propose what they call an Epistemic Separability Principle (ESP):

> ESP: Suppose that universe U contains within it a set of subsystems, S; such that every agent in an internally qualitatively identical state to agent A is located in some subsystem that is an

> element of S. The probability that A ought to assign to being located in a particular subsystem, $X \in S$, given that they are in U, is identical in any possible universe that also contains subsystems S in the same exact states (and does not contain any copies of the agent in an internally qualitatively identical state that are not located in S).
>
> <div align="right">(*op.cit.,* p. 16)</div>

Critics of Sebens' and Carroll's argument might well point up the 'ought' in the ESP and argue that the principle has simply been assumed in order to yield the desired result. However, I have argued that the ESP is entailed by the unitary interpretation of mind because it reinterprets the mentality of copies of agents in 'internally qualitatively identical' states (Tappenden 2017 §2). According to that reinterpretation blindfolded Alice does *not* split on making her quantum measurement. Rather, her mind spans the *up* and *down* branches. Her brain evolves into a superposition of two brains, one on each branch, but those brains do not instance the minds of distinct subjects because they are cognitively identical even though physically different in some respects. As a result, the measuring device in blindfolded Alice's environment post-measurement is a superposition of devices pointing *up* and *down* (*ibid.*, p. 14).

Given this reinterpretation of the setup it could seem that the incoherence problem for fission reappears since blindfolded Alice post-measurement is no longer *uncertain* about which branch she's on. Rather, if well-informed, she's certain that she's in the presence of a measuring device in superposition. However, as I've pointed out, Vaidmanian uncertaintly is easily recovered (*ibid.*). Let Alice remain blindfolded post-measurement and let a bell be rung on the *up* branch and a whistle blown on the *down* branch without Alice knowing which sound goes with which outcome. According to the unitary interpretation of mind she will then split into Alice$_{up}$ and Alice$_{down}$, each uncertain as to which branch she's on. The previous argument in response to Albert therefore still applies. The possibility of post-measurement, pre-observation uncertainty in fission contexts is all that's required to motivate rational action pre-measurement.

But the reappearance of the unitary interpretation of mind is noteworthy. It seems to put Sebens' and Carroll's justification of the Born-Vaidman rule on a less arbitrary, better motivated footing than their ESP. And a similar argument may be applied to (McQueen and Vaidman 2019). McQueen's and Vaidman's approach appeals to what they call the local supervenience principle:

> whatever happens in region A depends on the quantum description of this region and its immediate vicinity.
>
> <div align="right">(*ibid.*, §4)</div>

The idea is this. If blindfolded Alice performs her measurement in region A and then Bob performs a similar measurement in region B if and only if Alice's measuring device records *up*, then Alice's post-measurement, pre-observation subjective probability assignments should only take into account what has happened to *her* measuring device *not* what has or has not happened to Bob's device.

McQueen's and Vaidman's local supervenience principle can also be shown to be entailed by the unitary interpretation of mind because relative to blindfolded Alice post-measurement her local measuring device in region A becomes a superposition and Bob's device in region B either does or does not become a superposition depending on whether he initiates a measurement. Either way, when it comes to blindfolded Alice making her post-measurement subjective probability assignments it's *only* the state of her local superposed measuring device which she needs to take into account (which has evolved into a superposition of a device on the *up* branch, weight 0.7 and a device on the *down* branch, weight 0.3). When Alice's blindfold is removed post measurement she splits. Alice$_{up}$ is on the branch where her measuring device shows *up* and distant Bob's device is in superposition. And Alice$_{down}$ is on the branch where her measuring device shows *down* and distant Bob's device has remained in the ready state. And post-measurement, before Alice's blindfold is removed, Vaidman's post-measurement, pre-observation self-location uncertainty can be recovered with the help of a bell and a whistle linked to the device in region A.

The upshot is that if the Born-Vaidman rule is taken to be justified then an alternative response to the incoherence problem for the fission interpretation of branching is in place. Given the assumption that pre-measurement uncertainty *doesn't* apply in fission contexts, pre-measurement behavior can be guided by the *possibility* of post-measurement, pre-observation uncertainty. That argument is not widely recognized and deserves to be underlined, however, what's it got to do with the no-collapse postulate?

**2.7 Indefiniteness and uncertainty**

Both the Sebens-Carroll and McQueen-Vaidman arguments claim to show that blindfolded Alice$_{up}$ and Alice$_{down}$ should assign subjective probabilities to future observations (when the blindfolds are removed) equal to what they take the branch weights to be. Neither argument claims that branch weight is to be identified with objective probability but their reinterpretation in the light of the unitary interpretation of mind changes the perspective.

Recall that the unitary interpretation of mind entails that blindfolded Alice does not split post-measurement and Vaidmanian self-location uncertainty is recovered by causing Alice to split via use of the bell and whistle. So we have the conceptual juxtaposition of a single blindfolded Alice being in the presence of a measuring device in an indefinite pointer state and then Alice$_{up}$ and Alice$_{down}$ each being uncertain as to which branch they're on.

Dwelling on this juxtaposition is instructive. The sounding of the bell and whistle causes Alice to cease to be a subject who is certain, we can suppose, that she's in the presence of a measuring device in an indefinite state. She fissions into two subjects each being uncertain as to which branch they're on. In that case, is it really plausible that Alice post-measurement, pre-bell-and-whistle has no grounds for assigning subjective probabilities to the observation of *up* and *down* once the blindfold is removed whereas Alice$_{up}$ and Alice$_{down}$ *do* have grounds for subjective probability assignments? I suggest that such a radical change in point of view brought about by the bell-and-whistle is not plausible. There *are* grounds for post-measurement, pre-fission Alice to assign subjective probabilities in just the way Alice$_{up}$ and Alice$_{down}$ can and those grounds do *not* involve being in a state of self-location ignorance. They involve invoking the no-collapse postulate plus PP. The no-collapse postulate entails that the absolute squares of amplitude of the *up* and *down* elements of the superposed measuring device in blindfolded Alice's environment post-measurement, pre-fission are their objective probabilities relative to the ready state of the apparatus. And Alice, we can suppose, knows that when the blindfold is removed she'll split because she'll be observationally exposed to a measuring device in an indefinite pointer state. Applying PP, she should thus assign subjective probabilities of 0.7 and 0.3 to the future observation of *up* and *down* respectively.

The arguments so far have in a sense come full circle. I earlier argued that if the unitary interpretation of mind is coherent then it's coherent that Alice pre-measurement, knowing that she'll split, is in a state of uncertainty because the no-collapse postulate plus PP entail that she assigns subjective probabilities of 0.7 and 0.3 to the future observation of *up* and *down*. I then argued that even if pre-measurement uncertainty is set aside the incoherence problem can be resolved by showing that rational action pre-measurement can be assured via the possibility of post-measurement, pre-observation uncertainty, itself justified by the Sebens-Carroll and McQueen-Vaidman arguments whose key assumptions are entailed by the unitary interpretation of mind. Furthermore, the unitary interpretation of mind appears to have a fundamental role to play if stage theory is to be used to describe the partially ordered histories of fissioning subjects.

Clearly, trying to make sense of the no-collapse postulate in the context of Everett's fission interpretation of branching flirts with some radical metaphysics. Might there be a more conservative way to make sense of Everettian probability? Saunders, Wallace and Alastair Wilson think so since they all defend, in different ways, what Wallace calls the *Conservative View* of Everettian theory (*op.cit.*, p. 270). The idea is to replace the fission interpretation of branching with varieties of a partitioning linear histories interpretation. I shall summarize the latter and then suggest that it's not, as claimed, a deterministic theory since stochasticity appears to be covertly re-introduced.

**3 Overlap and divergence**

I should begin by noting that Saunders, Wallace and Wilson are sympathetic to the idea that objective probability should be identified with relative branch weight (Saunders 2010, p. 182; Wallace 2012, p. 141; Wilson 2013, pp. 771-72). The idea can seem to be suggested by the Deutsch-Wallace argument, as explained below. In that case, since the no-collapse postulate is the hypothesis that objective probability *is* relative branch weight, the Deutsch-Wallace argument, if good, would then become a justification of PP because it shows that subjective probabilities should be assigned equal to objective probabilities.

The *overlap* interpretation of branching was introduced by Saunders and Wallace in order to establish pre-measurement uncertainty for Everettian theory; their joint paper was entitled *Branching and uncertainty*. Instead of interpreting subjects as splitting, and so having dendritic histories, they interpret them as having conventional linear histories. They write:

> To conclude: if – as Lewis proposes – in cases of personal branching we say that there are two persons present even before the branch, it is at least somewhat natural to attribute two sets of thoughts to those persons; in the case of worlds branching, it becomes entirely natural. As a result, talk of uncertainty in the face of branching comes out as true.
> (Saunders and Wallace 2008, p. 303)

The reference is to David Lewis's analysis of thought experiments involving personal fission (Lewis, 1976). Applied to our model spin measurement, Alice$_{up}$ and Alice$_{down}$ are each to be *identified* with a linear history and those histories overlap prior to measurement, which is to say that they have their temporal parts in common. Saunders and Wallace refer to Lewis's analogy of overlapping roads (*ibid.*, p. 295). The road from Southville to Westville and road from Southville to Eastville may overlap between Southville and Northville, before they go their separate ways:

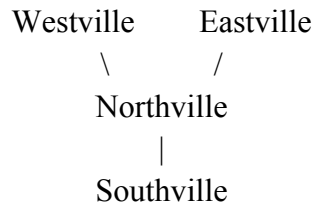

So Alice$_{up}$ and Alice$_{down}$ both exist before the measurement, but have their temporal parts in common. Saunders and Wallace differ from Lewis in concluding from this that 'it is at least somewhat natural to attribute two sets of thoughts to those persons' (*ibid.*, p. 303). In applying the idea to Everettian theory they extend this view of personal identity to the identity of worlds:

> As goes the incoherence problem of EQM [Everettian quantum mechanics], it is now rather clear, from Section 2, of what we are ignorant: we don't know which world—which branch, big-bang to end-of-time—is ours. It is lack of knowledge *de se*, uncertainty of where we are located, not as a stage $S$ but as a world-stage $<W,S>$ or world-time $<W,t>$, among the branching worlds.
> (*ibid.*, p. 301)

So, prior to measurement Alice is one of a host of subjects who have isomorphic bodies, each inhabiting one of a set of 'big-bang to end-of-time worlds' which, because they overlap, have remained isomorphic up until the measurement event, at which time the set partitions into a subset where the outcome *up* occurs and a subset where the outcome *down* occurs. In each world the outcome of the 'parallel' measurements are determined to be exclusively either *up* or *down* but it is in principle impossible for a subject to be able to predict what the outcome in her world will be. So prior to measurement Alice is necessarily uncertain as to which type of world she inhabits, one where *up* is destined to occur or one where *down* is destined to occur. She doesn't know whether she's an Alice$_{up}$ or an Alice$_{down}$. From her point of view, the measuring device in her environment behaves exactly *as if* the outcome were determined stochastically in the sense that *it will show only one outcome which is in principle impossible to predict and a subjective probability can rationally be assigned to the future observation of each possible outcome.*

Lewis's analysis of personal fission, which Saunders and Wallace modify in an attempt to generate pre-measurement uncertainty, depends on a metaphysics of persistence which Sider calls the *worm view* (1996, p. 433). In contrast to stage theory, which identifies subjects and objects with the momentary temporal parts of their histories, worm theory identifies subjects and objects with linear histories which manifest themselves as 'spacetime worms'.

The problem presented by personal fission was that Alice$_{up}$ and Alice$_{down}$ must be regarded as numerically distinct subjects and there's no criterion for supposing that either one was Alice. The stage theory solution to that problem counts Alice, Alice$_{up}$ and Alice$_{down}$ as each a distinct subject and introduces the temporal counterpart relation to describe persistence. The worm theory solution proposed in (Lewis 1976) takes Alice$_{up}$ and Alice$_{down}$ to be identified with linear histories which have their temporal parts in common prior to fission. According to worm theory personal fission does not involve the splitting of a subject, it involves the partitioning of the numerically distinct linear histories which *are* the persons involved. Note that histories which partially overlap are numerically distinct despite having temporal parts which are numerically identical.

The question then arises as to how each of the Alices pre-measurement can assign subjective probabilities to either being an Alice$_{up}$ or an Alice$_{down}$. Deutsch was the first to address this problem, which likewise arose for his proposal to supplement quantum theory with what he called Axiom 8:

> The world consists of a continuously infinite-measured set of universes.
> By a 'measured set' I mean a set together with a measure on that set.
> (1985 p. 20)[4].

Whereas Saunders' and Wallace's partitioning worlds are based on decoherent history theory Deutsch proposes an infinite set of initially isomorphic worlds which partitions exactly *as if* stochastic processes were taking place in each of them even though processes in each world are *assumed* to be deterministic. Deutsch's Alices need to assign subjective probabilities equal to the subset measures which correspond to branch weights for his scheme, and in order to justify that he introduced a decision-theoretic argument, later elaborated by Wallace. (Deutsch 1999, Wallace 2012, pp. 160-189). The Deutsch-Wallace argument claims to prove that Alice should assign subjective probabilities to the future observation of *up* and the future observation of *down* equal to the weights of the *up* and *down* branches. Unlike the Sebens-Carroll and McQueen-Vaidman arguments, which consider the perspective of the post-measurement, pre-observation subject, the Deutsch-Wallace argument considers the perspective of the pre-measurement subject. All three arguments claim to show that a subject's assigned subjective probabilities to future observations should be equal to what s/he assumes the relative branch weights of the outcomes to be. So all three arguments, if good, amount to a justification of PP on the assumption that relative branch weight *is* objective probability.

---

[4] An idea developed in a different way in (Barrett 1999, pp. 179-84).

On the linear histories analysis any subject's measuring device behaves exactly *as if* it were detecting a stochastic process, the reason being that it is destined so show just *one* outcome which cannot *in principle* be predicted but for which *probabilities* can be assigned. If *branch weight* acts as if it were objective probability then there's no reason not to *identify* it with objective probability: that 'functionalist' argument has been endorsed by Saunders and Wallace (Saunders 2010, p. 182; Wallace 2012, p. 141). By the *same* argument, if Alice's detector behaves *as if* it were measuring a stochastic process then there's no reason *not* to think that it *is* detecting a stochastic process.

So the Saunders-Wallace linear histories proposal is arguably *not* a deterministic theory; it's a stochastic many-worlds theory in disguise. The supposed determinism in each big-bang-to-end-of-time world is not *physical* it arises out of an alternative *metaphysical* interpretation of the branching process.

## 3.1 Divergence

Subsequent to the overlap proposal, Saunders has suggested another way of conceiving of branching as the partitioning of linear histories (Saunders 2010, p. 196). To contrast this concept of partitioning to that involving overlap Saunders uses the term 'divergence', following (Lewis 1986, p. 206). He writes:

> worlds in EQM [Everettian quantum mechanics] do not diverge in
> the sense of being physically disconnected (they are not
> physically disconnected, because they superpose, but the issue is
> whether or not they overlap)
>
> (*op.cit.*, p. 197)

Saunders argues that the Heisenberg picture suggests that the pre-branching segments of a set of 'big-bang-to-end-of-time worlds' can be thought of as being numerically distinct rather than overlapping, and that the mathematics of quantum theory doesn't distinguish between overlap and divergence. The implication is that we are free to think of them as numerically distinct; as separate in the sense of being 'superposed'. Wilson has developed the idea into modified form of Lewis's 'modal realism' (Wilson 2013; forthcoming).

Wallace disagrees with Saunders and Wilson, maintaining that there's no important difference between the overlap and divergence versions of the partitioning linear histories interpretation of branching (*op.cit.*, pp. 286-7). But the divergence of views here is of no great consequence; the aim of all is to establish pre-measurement uncertainty, thought to be absent for a subject who splits. As Wallace has put it:
:
> none of this is to concede that we cannot make sense of uncertainty, or of alternative possibilities, in the Everett interpretation. In fact, we can make sense of them just fine, as Chapter 7 will argue.
>
> (*op.cit.*, p. 119)

In Chapter 7 he reiterates the ideas in (Saunders and Wallace 2008).

And both the overlap and divergence versions of partitioning linear histories, whether or not there's a significant difference between them, view quantum processes *within* a big-ban-to-end-of-time world *as if* they were stochastic. So, again, by the functionalist argument the branch weights of partitioning linear histories arise out of stochastic processes, not a deterministic process. It seems that only the fission interpretation of branching can yield a *deterministic* quantum mechanics, if anything can.

And, given the unitary mind interpretation of fission, the Deutsch-Wallace argument applies alongside the Sebens-Carroll and McQueen-Vaidman arguments regarding a post-measurement, pre-observation subject. The reason being that Alice post-measurement, pre-observation is in the presence of a superposition in just the same way that Alice pre-measurement was. All that has happened during the measurement process is that the object to be measured, which is a microscopic superposition, has amplified via decoherence so as to put Alice's pointer in a superposition of pointing *up* and *down*. From the unitary mind point of view, what distinguishes the Sebens-Carroll and McQueen-Vaidman arguments from the Deutsch-Wallace argument post-measurement, pre-observation is that the former need the bell and whistle whereas the latter doesn't.

Finally, to underline the point, all three of these 'derivations of the Born rule' become possible justifications of PP if the no-collapse postulate is accepted. Because they aim to demonstrate that subjective probabilities for future observations should equal what the objective probabilities (branch weights) are taken to be.[5]

---

[5] Further thoughts on overlap and divergence can be found in (Tappenden 2019).

## 4 Damned lies and statistics.

An objection sometimes raised against Everettian theory is that it makes quantum mechanics unfalsifiable because all 'possible' outcomes of measurement processes actually occur. That point is forcibly made by Emily Adlam (2014). In this section I shall consider how that problem looks given the no-collapse postulate.

Adlam favorably cites Albert (*ibid.*, p. 26). Here's an expanded version of that citation:

> What needs to be looked into, in order to answer the question of whether to believe the fission hypothesis is correct, is whether or not the truth of that hypothesis is explanatory of our empirical experience. And that experience is of certain particular sorts of experiments having certain particular sorts of outcomes with certain particular sorts of frequencies—and not with others. And the fission hypothesis (since it is committed to the claim that all such experiments have all possible outcomes with all possible frequencies) is *structurally incapable* of explaining anything like that.
>
> (Albert 2010, p. 359, original emphasis)

It's notable that Adlam, in extracting her citation from the above passage, replaces reference to what Albert calls 'the fission hypothesis' with the phrase 'Everettian quantum mechanics'. But that appellation, often referred to as EQM, has come into general currency not least via the writings of Saunders, Wallace and Wilson who specifically argue for partitioning linear histories interpretations of branching rather than the concept of splitting (fission) which was originally introduced by Everett!

But Albert has left something out. Our experience of certain particular sorts of experiments having certain particular sorts of outcomes with certain particular sorts of frequencies does not license our belief that quantum mechanics is correct without a further assumption. That assumption is that it's highly improbable that the world has conspired to give us results to our experiments which suggest that quantum mechanics is true when in fact it's false. In other words, we have to assume that the data we've collected is a reliable guide to the way the world is, that it's a fair sample not a freak sample. And of course physicists are *obliged* to assume that a sufficiently long run of experiments *does* give a reliable result; it would be irrational to do otherwise. And that goes for *any* physicists anywhere in Everett's multiverse. So physicists in 'maverick' branches, where frequencies of outcomes do *not* confirm quantum mechanics, are obliged to believe that they have disconfirming evidence. It would be as irrational for them to believe they inhabit low-probability branches as it would be for a stochastic theorist to believe that an errant experimental run was due to getting an improbable sample.

Adlam precedes her quote with this thought:

> Thus even if we do happen to occupy one of those branches in which the relative frequencies are close to the mod-squared amplitudes, this is purely a matter of good luck and not a fact for which the amplitudes bear any responsibility, so the amplitudes cannot possibly be responsible for our having made the observations that we have.
>
> <div align="right">(*op.cit.*)</div>

Good luck? If Adlam is implying that it's *improbable* that we should get evidence confirming quantum mechanics that is clearly not the case given the no-collapse postulate if, in fact, quantum mechanics is correct. For if it *is* correct then the branches with the right frequencies for long experimental runs have by far the highest objective probability. And in assuming that we get fair samples, as we're obliged to do, we assume that the branches we inhabit have high objective probability and so conclude that quantum mechanics is correct. Of course, any physicist launching into an experiment to test quantum mechanics today will be a physicist getting disconfirming evidence on some branches, given the stage theory analysis of fission, because there's always some probability of getting disconfirming evidence even if the theory is true. But on the assumption that quantum mechanics *is* correct the branches on which that occurs will be of very low objective probability and should thereby be accorded very low credence in advance. Just like the physicist who believes that quantum processes are stochastic, the physicist who believes that they are dendritic must accord some credence to coming to find quantum mechanics disconfirmed. Both stochastic and dendritic theorists have reason to run experiments.

**5 The metaphysical consequences of the unitary interpretation of mind**

I have argued that the no-collapse postulate and PP, combined with the fission interpretation of branching, can yield a deterministic and objectively probabilistic quantum mechanics. The unitary interpretation of mind appears to have a fundamental role to play in making that combination of ideas coherent in the face of contrary intuitions. The implication is that Everettian theory requires a revisionary metaphysics for the constitution of objects in our environment, as I shall now explain.

    The issue is tied up with an argument which originated in a seminal thought experiment introduced by Hilary Putnam (1975). That led Putnam to inveigh against what he called metaphysical realism, the idea that an objective 'external word' exists independently of mentality. His argument is succinctly put in (Putnam 1981, Ch. 1). It derives from the concept of semantic externalism which was further refined by Tyler Burge, inspired by Putnam's original thought experiment (Burge 1979; 1982). Burge described this change of perspective in analytic philosophy as a move away from:

> the elderly Cartesian tradition [which puts] the spotlight on what exists or transpires 'in' the individual – his secret cogitations, his innate cognitive structures, his private perceptions and

> introspections, his grasping of ideas, concepts or forms. […] This [anti-Cartesian] tradition has dominated the continent since Hegel. But it has found echoes in English-speaking philosophy during this [20th] century in the form of a concentration on language. Much philosophical work on language and mind has been in the interests of Cartesian or behaviorist viewpoints that I shall term 'individualistic'. But many of Wittgenstein's remarks about mental representation point up a social orientation that is discernable from his flirtations with behaviourism.
>
> (Burge 1979, p. 73)

Putnam also refers to Ludwig Wittgenstein as a harbinger of the 'anti-individualistic' perspective in analytic philosophy (Putnam 1981, pp. 3, 7 and 20-21).

The unitary interpretation of mind is intimately connected with the Putnam-Burge 'Twin Earth' arguments for anti-individualistic semantic externalism, as has been pointed out (Tappenden, 2017, p. 15). The best way to demonstrate that is to consider Putnam's original thought experiment. I shall give a slightly alternative presentation of the idea to capture the essence of the arguments as succinctly as possible. We are to imagine that when it was 1750 on Earth there existed somewhere in the universe a Twin Earth which was isomorphic to Earth except that every occurrence of water on Earth was matched by an occurrence of twater, a different thirst-quenching clear liquid, on Twin Earth. Twin Earth might exist anywhere in spacetime, there's no requirement of simultaneity. Whereas water is $H_2O$, twater is XYZ and water and twater would have been indistinguishable for anyone living in on Earth in 1750, before modern chemistry, supposing that a sample of twater were made available to Earthlings then. Furthermore, implausible as it may be, the molecular difference between water and twater would not have had any noticeably different consequences for the environments on the two planets.

The externalist argument is that when a person on Earth in 1750 thought about the stuff to which they referred by using the term 'water' what they were thinking about was $H_2O$. And when that person on Earth was thinking about water it is possible that there was a person with an isomorphic body on Twin Earth who was thinking about XYZ (we have to ignore the fact that people on Earth have water in their bodies and people on Twin Earth have twater, or just suppose that that's not cognitively significant). Thinking about water is different from thinking about twater for they are thoughts about different stuffs. The thoughts have different semantic contents despite the fact that the thinkers on Earth and Twin Earth have isomorphic bodies. So the semantic contents of at least some thoughts cannot be wholly determined by what goes on in thinkers' bodies; that's semantic externalism. As Putnam put it, '*meanings just aren't in the head*' (Putnam 1981, p. 19, original emphasis).

Juhani Yli-Vakkuri has recently made explicit one of the concepts on which such arguments depend: a relation of 'being corresponding beliefs of duplicate subjects' (Yli-Vakkuri,2018, p. 85) where: 'Beliefs here must be thought of as tokens rather than types' (*ibid.*, p. 83).

Despite its intuitive appeal, the idea that isomorphic doppelgangers are 'duplicate subjects' having distinct 'corresponding' token beliefs is exactly what is challenged by the unitary interpretation of mind which entails that the two matched doppelgangers on Earth and Twin Earth share a single token thought when thinking about the stuff in matched glasses of the clear liquids. That token thought refers to a single sample which is the *set* of the sample of $H_2O$ and the sample of XYZ. Any Twin Earth argument for semantic externalism is thereby countered since there is a single subject referring to a single object rather than two subjects with isomorphic bodies referring to two distinct objects with different constitutions.

This counterargument requires assuming that any set of environmental objects is an object which has all and only the properties which its elements share, with some exceptions to be listed later. So the single sample referred to in 1750 has a molecular constitution which is indefinite because it is a set whose elements have molecular constitutions which are different. Introducing the concept of concrete objects which have indefinite properties to an argument against semantic externalism establishes a link with quantum mechanics.

The link with quantum fission emerges if we think about our relation to our ancestors in 1750. If Twin Earth really could exist then it must be possible that in 1750 our Earthly ancestors had minds which spanned the two planets. In which case the Thames in William Hogarth's London did not necessarily contain water, it could have contained a strange mixture of $H_2O$ and XYZ which lacked a definite molecular constitution. And what would have happened when chemists first managed to probe the constitution of the stuff in rivers and rain is that they split into chemists finding $H_2O$ here on Earth and chemists finding XYZ on what would now be faraway Twin Earth.

This point connects with Sarah Sawyer's recent thoughts on semantic externalism. She argues that it guarantees the stability of the 'subject matter' of concepts such as that of water (Sawyer 2018, §4). In other words it guarantees that what Hogarth called water is the very same stuff that we call water now. If she's right, the only possible defense of semantic *internalism* requires that it is possible, given the possibility of Twin Earth, that the stuff in Hogarth's Thames was *not* $H_2O$, which is what follows from the unitary interpretation of mind.

If semantic internalism is correct then mental content is exclusively a property of localized physical objects, most plausibly brains or subsystems of brains; it's not something which involves a relation between brains, their bodies and their environments. An accidental cerebral episode (ACE) which is isomorphic to a normally embodied and environmentally embedded brain bears all the same mental content. An ACE may pop into existence in deep space, or in the heart of a star, due to quantum fluctuations in what Bertrand Russell once called 'a climax of improbability' (Russell 1954, p. 33). In recent years ACEs in space have come to be known as Boltzmann brains and their widespread existence in the vast tracts of spacetime in which we live is taken by many cosmologists to be inevitable. And the widespread existence of ACEs is certainly inevitable if quantum fluctuations are dendritic rather than stochastic. That vividly raises a problem of Cartesian skepticism: is your brain, now, normally embodied and embedded in the way you think it is or is it an ACE? The unitary interpretation of mind brings a new perspective to that old problem.

So here's that semantic internalist perspective in a nutshell. To identify an individual mind with an individual brain is a mistake similar to identifying a sentence with an inscription of it. An individual mind supervenes on any *set* of sufficiently isomorphic brains and/or functionally equivalent objects such as, perhaps, complex computers. In any sort of situation where there are multiple doppelgangers in isomorphic environments but where the quantity of 'worlds' may be any number N, the mind of a subject is instanced by a set of N brain-like objects. Token thoughts and utterances are instanced by sets of N cerebral and sonic objects. Any environmental object to which a subject is able to indexically refer, including their body and brain, is constituted by a set of N isomorphic objects (note that elements of such sets are not contributed by the environments of ACEs). If there happens to be, in fact, only a single instance of a subject's brain and environment, then environmental objects are taken to be self-membered singleton sets known as Quine atoms (Tappenden 2017, p. 10). To avoid Russellian set-theoretic paradox self-membership is restricted to Quine atoms. Note that any *aggregate* in the environment would be a Quine atom, as would be its parts.

The unitary interpretation of mind entails that any set of environmental objects has all and only the properties which its elements have in common, with exceptions for self-membership, number of elements, value-definiteness and mental properties. Mental properties are excluded just because subjects are individuated by their mental contents, not the instances of those contents. As a consequence, the set of the two gloves making an ordinary pair is a glove which has the mass of one and the property of handedness but is itself neither a left hand nor a right hand glove. To see why, consider two identical rooms with matched doppelgangers in them. There will be a single subject whose mind spans the two rooms; call her Diana. Now take an ordinary pair of gloves, put each glove in one of a pair of type-identical boxes and put each of the boxes in corresponding locations in each of the two rooms.

Diana sees a single box which is the set of the two boxes. If she weighs it the doppelgangers move in concert to place the boxes on matched scales which each show the weight of one box plus one glove, so Diana perceives her box as containing an object with the weight of one glove. That object is the set of the two gloves. It has the property of handedness because the two gloves have that property in common, but Diana's glove is neither left handed nor right handed because the gloves are neither both left handed nor both right handed. If Diana opens her box she fissions into a subject finding a left hand glove and a subject finding a right hand glove.

So the unitary interpretation of mind entails that we're surrounded by objects with indefinite properties. Any set of objects in the environment is a concrete object which only has definite properties to the extent that its elements have properties in common. And recall the Hydra view which suggests that the elements of a superposition are a novel type of non-spatial, non-temporal part. If the unitary interpretation of mind is correct then the elements of a superposition are not *parts* at all, they're elements *in the set-theoretic sense.* Schrödinger's cat, in the imaginary causally isolated interior of its box, is the set of a dead cat and a live cat whose masses and volumes remain much the same whilst their quantum amplitudes change.

Finally, here's an oddity which falls short of paradox. Consider *three* identical rooms with matched doppelgangers in them. Given the unitary interpretation of mind there's one subject whose mind spans the three rooms; call her Triana. Now introduce a sample of a different shade of grey in each room: Dark, Medium and Light, where there's a just noticeable difference between Dark and Light but not between Dark and Medium nor between Medium and Light. If Triana sets eyes on the set of samples she will fission into Diana$_{Dark}$ and Diana$_{Light}$ each of whose minds will be instanced by a set of two brains one of which will be common to both of them. How is that possible?[6]

On the currently not implausible hypothesis that minds are instanced by mechanisms it may well be that a particular mind can be instanced by any one of a range of mechanisms whose differences are not great enough to generate difference of perception. As soon as a difference becomes noticeable, whether consciously or unconsciously, the subject fissions. Diana$_{Dark}$ and Diana$_{Light}$ have a brain in common because that brain implements a mechanism compatible with both perceiving Dark and perceiving Light.

**6 In conclusion**

I've argued that the introduction of partitioning linear histories interpretations of branching to Everettian theory in order to introduce pre-measurement uncertainty is unnecessary and arguably involves covertly stochastic quantum processes. In which case only Everett's original fission interpretation of branching can yield a deterministic theory. If pre-measurement uncertainty is thought to be necessary for fission then that can be made coherent via the unitary interpretation of mind; if it's not thought to be necessary then rational action pre-measurement can still be justified via the Born-Vaidman rule and the Sebens-Carroll and McQueen-Vaidman arguments. However, those latter arguments themselves are more fundamentally motivated by adopting the unitary interpretation of mind, which may also be required if the history of a fissioning subject is to be understood via stage theory as a partially ordered series of subjects. Also, given the no-collapse postulate, the Deutsch-Wallace, Sebens-Carroll and McQueen-Vaidman arguments become attempts to justify PP.

So it appears that the unitary interpretation of mind has a key role to play, together with the no-collapse postulate and PP, in making Everettian probability coherent. In which case a revised concept of the metaphysical constitution of environmental objects is required. If that's right, Everettian theory does indeed usher in a 'Copernican' revolution, just as Everett envisaged (*op. cit.,* p. 460). Each of us is splitting incessantly because of decoherence processes; each of us has multiple futures. Though the details are a complex matter, many processes in nature, such as the weather, are known to be chaotic in the sense that they are very sensitive to small differences of input, the so-called butterfly effect. So if you're planning an outdoor birthday party months in advance in a temperate climate it's plausible that the combined weights of the branches where it will be fine are substantial, as well as the combined weights of the branches where it will rain. There is even rather detailed work which suggests that significant quantum fission takes place for an ordinary coin toss (Albrecht and Phillips 2014).

---

[6] My thanks to Oliver Pooley for raising this question.

Everett's thesis supervisor, John Archibald Wheeler, wrote the following about (Everett 1957):

> It is difficult to make clear how decisively the 'relative state' formulation drops classical concepts. One's initial unhappiness at this step can be matched but few times in history.
>
> (Wheeler, 1957, 464)

When Wheeler was asked, early in the 1990s, why he had given up on Everett's idea he replied, 'It's too philosophical'.[7]

**Acknowledgements**

I wish to thank Jeff Barrett, David Deutsch, Douglas Campbell, Andrew F. Knight, John Ponsonby, Douglas Porpora, Simon Saunders, Mauricio Suàrez and David Wallace for useful comments. And in particular two anonymous referees for detailed and searching critiques which led to considerable revision of the original submission.

---

[7] As reported to me by Simon Saunders, who posed the question.